\documentclass[11pt,twoside]{article}
\usepackage{asp2014}

\aspSuppressVolSlug
\resetcounters

\begin{document}

\title{The Very Fast Evolution of the VLTP Object V4334 Sgr}
\author{P.A.M. van Hoof$^1$, S. Kimeswenger$^2$, G.C. Van de Steene$^1$, A.A. Zijlstra$^3$, M.~Hajduk $^4$ and F. Herwig$^5$
\affil{$^1$Royal Observatory of Belgium, Brussels, Belgium; \email{p.vanhoof@oma.be, gsteene@oma.be}}
\affil{$^2$Universidad Cat\'olica del Norte, Antofagasta, Chile; \email{Stefan.Kimeswenger@gmail.com}}
\affil{$^3$Jodrell Bank Centre for Astrophysics, Manchester, United Kingdom; \email{a.zijlstra@manchester.ac.uk}}
\affil{$^4$Nicolaus Copernicus Astronomical Center, Torun, Poland; \email{Marcin.Hajduk@astri.uni.torun.pl}}
\affil{$^5$University of Victoria, Victoria, Canada; \email{fherwig@uvic.ca}}}

% This section is for ADS Processing.  There must be one line per author.
\paperauthor{P.A.M. van Hoof}{p.vanhoof@oma.be}{0000-0001-7490-0739}{Royal Observatory of Belgium}{Astronomy \& Astrophysics}{Brussels}{}{1180}{Belgium}
\paperauthor{S. Kimeswenger}{Stefan.Kimeswenger@gmail.com}{}{Universidad Cat\'olica del Norte}{Instituto de Astronom\'\i a}{Antofagasta}{}{}{Chile}
\paperauthor{G.C. Van de Steene}{gsteene@oma.be}{}{Royal Observatory of Belgium}{Astronomy \& Astrophysics}{Brussels}{}{1180}{Belgium}
\paperauthor{A.A. Zijlstra}{a.zijlstra@manchester.ac.uk}{}{Jodrell Bank Centre for Astrophysics}{}{Manchester}{}{M13 9PL}{United Kingdom}
\paperauthor{M. Hajduk}{Marcin.Hajduk@astri.uni.torun.pl}{}{Nicolaus Copernicus Astronomical Center}{}{Torun}{}{87-100}{Poland}
\paperauthor{F. Herwig}{fherwig@uvic.ca}{}{University of Victoria}{Department of Physics and Astronomy}{Victoria}{British Columbia}{V8P 5C2}{Canada}

\begin{abstract}
V4334 Sgr (Sakurai's object) is an enigmatic evolved star that underwent a
very late thermal pulse a few years before its discovery in 1996. It ejected a
new, hydrogen-deficient nebula in the process. Emission lines from the newly
ejected gas were first discovered in 1998 (He\,{\sc i} 10830~\AA) and 2001
(optical). We have monitored the optical emission spectrum since. From 2001
through 2007 the optical spectrum showed an exponential decline in flux,
consistent with a shock that occurred around 1998 and started cooling soon
after that. In this paper we show that since 2008 the line fluxes have been
continuously rising again. Our preliminary interpretation is that this
emission comes from a region close to the central star, and is excited by a
second shock. This shock may have been induced by an increase in the stellar
mass loss and wind velocity associated with a rise in the stellar temperature.
\end{abstract}

\section{Introduction}

V4334 Sgr (a.k.a.\ Sakurai's object) is the central star of an old planetary
nebula (PN) that underwent a very late thermal pulse (VLTP) a few years before
its discovery in 1996 \citep{Na96}. During the VLTP it ingested its remaining
hydrogen-rich envelope into the helium-burning shell and ejected the processed
material shortly afterwards to form a new, hydrogen-deficient nebula expanding
at a velocity of approximately 300~km\,s$^{-1}$ inside the old PN. The star
brightened considerably and became a very cool, born-again asymptotic giant
branch star with a spectrum resembling a carbon star. After a few years, dust
formation started in the new ejecta and the central star became highly
obscured. Emission lines were discovered: first He\,{\sc i} 10830~\AA\ in 1998
\citep{Ey99}, later in 2001 also optical forbidden lines from neutral and
singly ionized nitrogen, oxygen, and sulfur, as well as very weak H$\alpha$
\citep{Ke02}. The distance to V4334 Sgr is poorly known, but is likely around
3 -- 4~kpc.

\section{Evolutionary Models}

Sakurai's object baffled the scientific community with its very fast
evolution, must faster than pre-discovery models predicted. Three evolutionary
models have been proposed to explain the fast evolution, all focusing on the
hydrogen ingestion flash (HIF) in the helium burning shell. \citet{He01} and
\citet{LM03} assume that hydrogen burning takes place close to the stellar
surface due to the suppression of convection by the HIF. This is investigated
further by Herwig using full 3D hydro models \citep{He11, He14}. These models
show that the hydrogen ingestion proceeds through a global non-radial
instability, which facilitates the transition from one spherically symmetric
state into another one, and could never be computed in 1D models. \citet{LM03}
were the first to predict the double-loop evolution in the HR diagram, later
confirmed by Herwig's model in \citet{Ha05}. \citet{MB06} claim that they can
reproduce very fast evolution by using very small time steps, but without
changing the mixing physics. All of these models could be improved by
constraining them with the temporal evolution of the stellar temperature. This
was reasonably straightforward when the central star was still directly
observable, but is much more difficult now that the star is heavily obscured.

\articlefigure{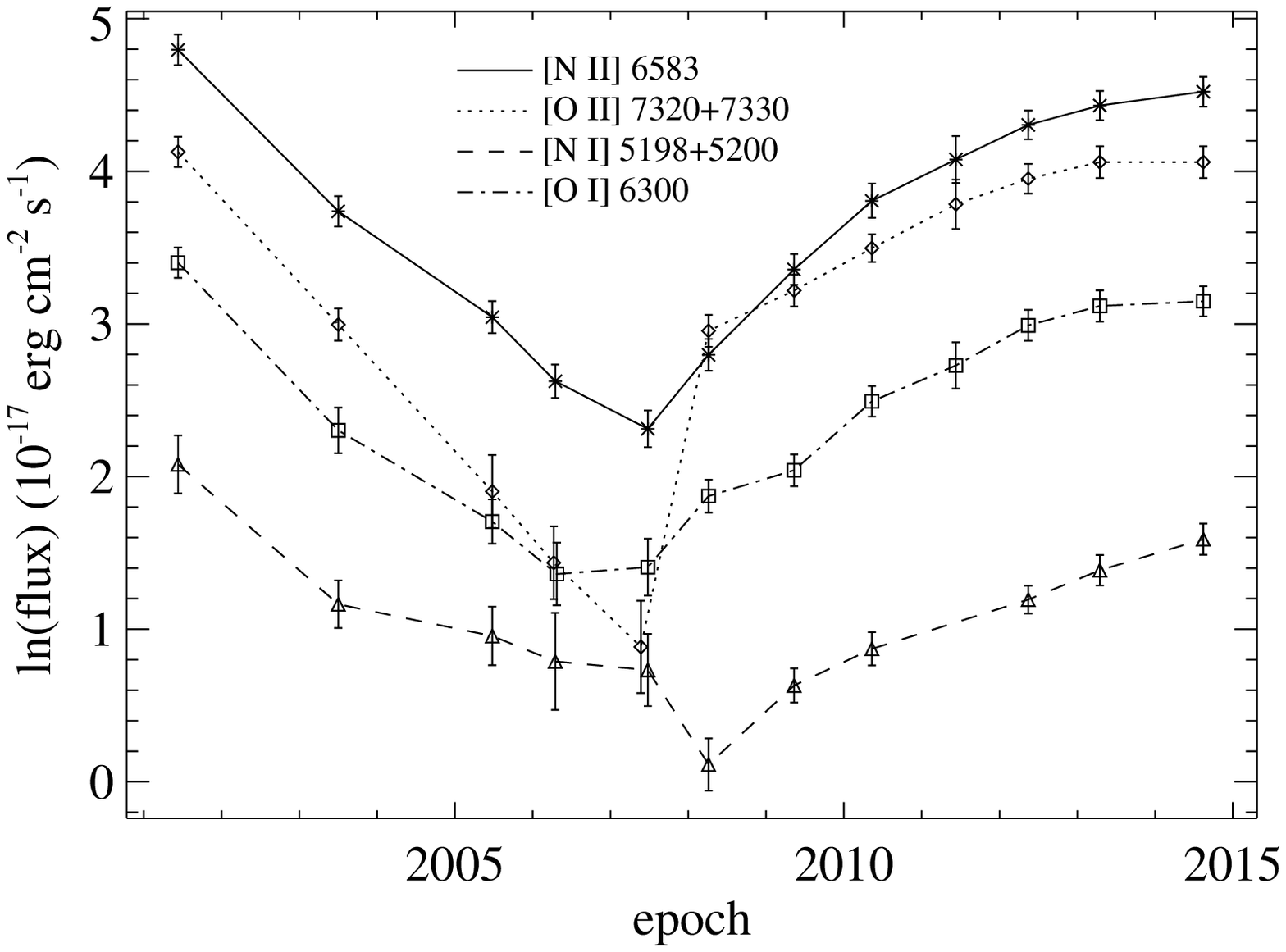}{evol}{The evolution of the flux of various
  selected lines as a function of time.}

\section{Optical Observations}

We have been monitoring the evolution of the optical emission line spectrum
since 2001 using low-resolution spectra taken with FORS1 and FORS2 on the
ESO-VLT. The goal of this monitoring program is to derive the central star
temperature as a function of time. First progress reports can be found in
\citet{vH07, vH08}.

The optical lines initially showed an exponential decline in intensity, and
also a decreasing level of excitation. This trend continued until 2007.
Between 2001 and 2007 the optical spectrum is consistent with a shock that
occurred before 2001, and started cooling and recombining afterwards. The low
electron temperature derived from the [N\,{\sc ii}] lines in 2001 (3200 --
5500~K) and the [C\,{\sc i}] lines in 2003 (2300 -- 4300 K) is consistent with
this \citep{vH07}. The earliest evidence for this shock is the detection of
the He\,{\sc i} 10830~\AA\ recombination line in 1998 \citep{Ey99}. This line
was absent in 1997. The shock must have occurred around 1998 and must have
stopped soon after, leaving cooling and recombining gas in its wake.

\begin{figure}[!ht]
\centering \leavevmode
\includegraphics[width=.95\textwidth]{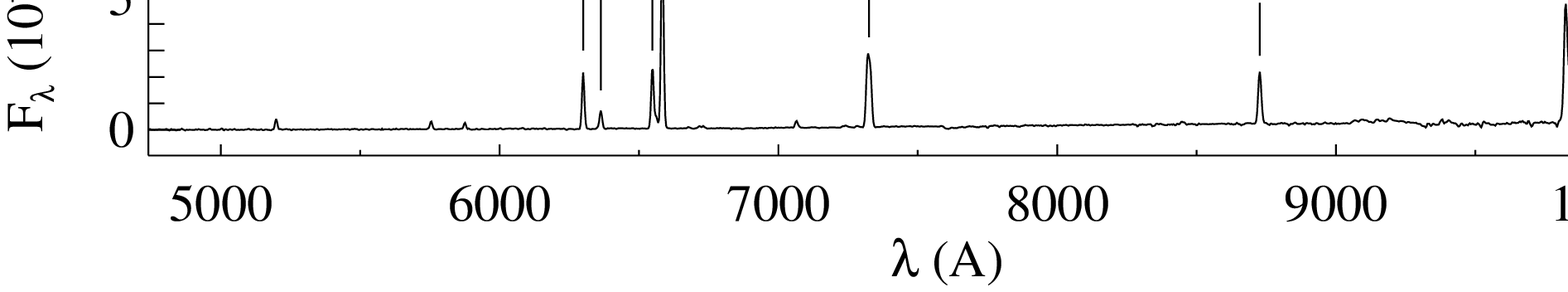}{}
\includegraphics[width=.95\textwidth]{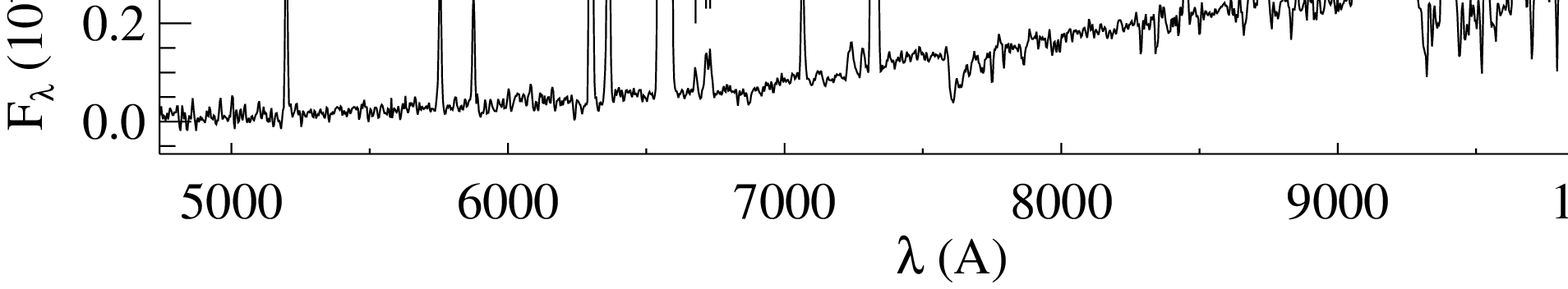}{}
\caption{The FORS2 spectrum observed in 2013. The top panel shows the complete
  spectrum, and the bottom panel shows a zoom-in on the weaker lines in the
  same spectrum.\label{spec}}
\end{figure}

All line fluxes have been increasing since 2008! This is shown in
Fig.~\ref{evol} for some selected strong lines. The spectrum we observed in
2013 is shown in Fig.~\ref{spec}. This confirms the trend for the [C\,{\sc i}]
9824 and 9850~\AA\ doublet reported by \citet{HJ14}. There are two exceptions:
[O\,{\sc i}] 6300~\AA\ already started increasing in 2007 and the [N\,{\sc i}]
5198 and 5200~\AA\ doublet still decreased in flux in 2008. However, these
exceptions may not be real as the lines in question suffer from strong
telluric contamination. Also note that there is a strong discontinuous jump in
the [O\,{\sc ii}] flux in 2008.

Here we report the first detection of several helium lines in 2008 (He\,{\sc
  i} 5876 and 7065~\AA) and 2009 (He\,{\sc i} 6678~\AA). Since 2010 the
[S\,{\sc ii}] 6716 and 6731~\AA\ doublet is detected again. This is the first
time this doublet is seen since 2001. This is due to a better signal-to-noise
ratio in the spectra and the fact that the lines are brightening again.
\citet{Ke02} detected H$\alpha$ at 5\% of the strength of [N\,{\sc ii}]
6583~\AA. Our spectra confirm this detection with a slightly greater strength
of 7\% of the [N\,{\sc ii}] line.

\section{Discussion}

\citet{HJ14} believe that the reheating of the central star has started,
mainly based on an increase in the blackbody temperature of the dust. This
suggests that an onset of photoionization could be the cause of the rising
line fluxes. However, in the most recent VLA observations V4334 Sgr was only
barely detected at around 50~$\mu$Jy in 2012 and around 150~$\mu$Jy in 2013,
indicating that the radio flux must have dropped since the detections
presented in \citet{Ha05} and \citet{vH07, vH08}. The last value they reported
was 550~$\mu$Jy in 2007. This seems inconsistent with an onset of
photoionization. Alternatively, the sudden jump in the [O\,{\sc ii}] flux in
2008 could point to a second shock as the cause of the rising fluxes and we
will adopt this as our working hypothesis.

\citet{Ke02} observed two components in the shock-excited [N\,{\sc ii}] 6548
and 6583~\AA\ lines, one at a radial velocity of $-350$~km\,s$^{-1}$ and one
at $+200$~km\,s$^{-1}$ relative to the central star. The blueshifted component
was much stronger, resulting in the fact that the unresolved lines appeared
blueshifted in the FORS spectra at about $-270$~km\,s$^{-1}$ relative to the
central star. However, since 2011 we see a clear shift in radial velocity and
the emission lines now appear at about $-120$~km\,s$^{-1}$ relative to the
central star. This indicates that the emission lines seen up to 2007 came from
a different region than those seen later on. Doing spectro-astrometry on the
recent spectra indicates that the emission comes from a region less than
100~mas in size in the EW direction. \citet{HJ14} find a larger extent in the
(roughly) NS direction. This suggests that the bipolar structure seen by
\citet{Ch09} and \citet{HJ14} could be the origin of the shock emission.

\articlefiguretwo{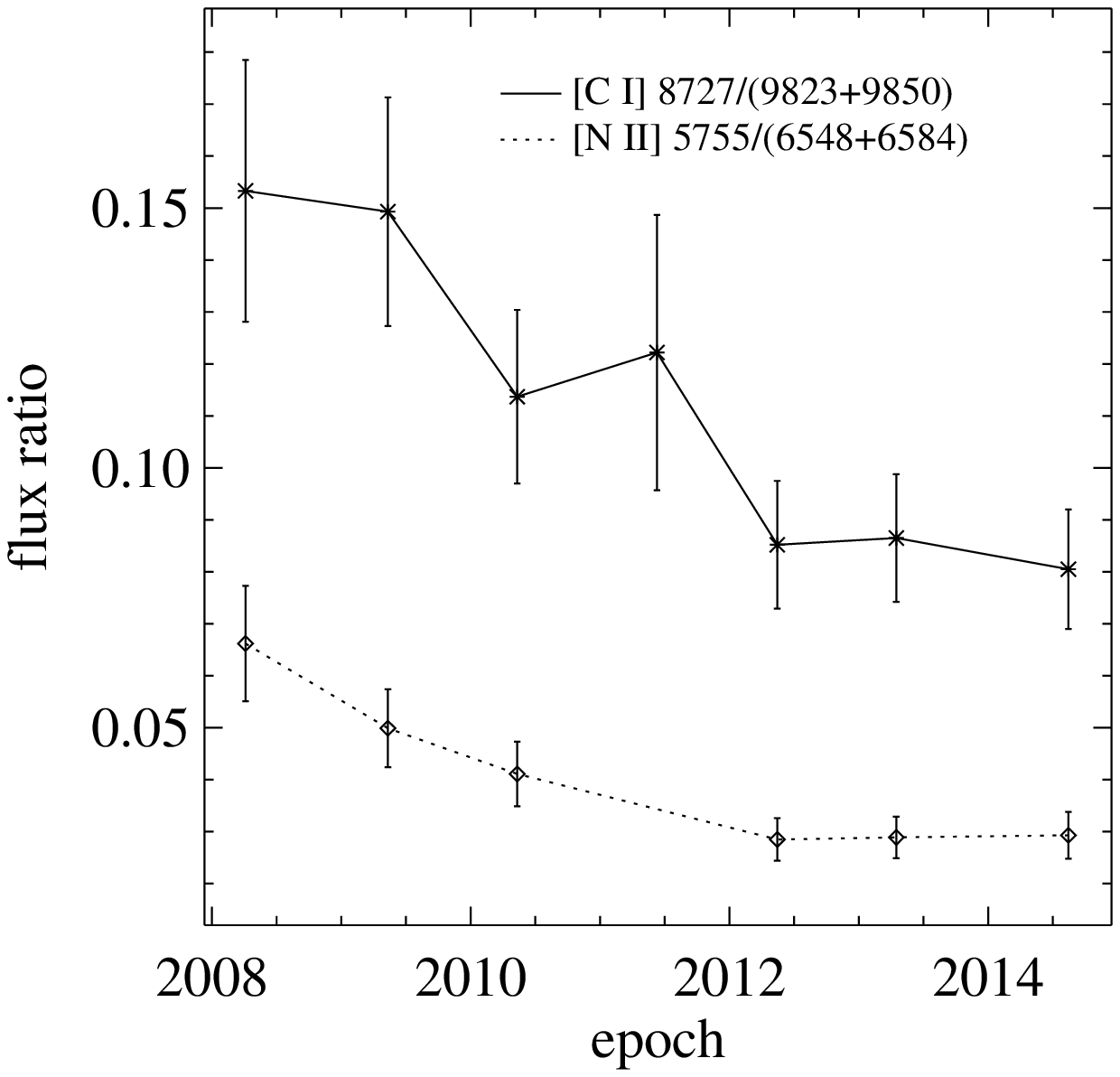}{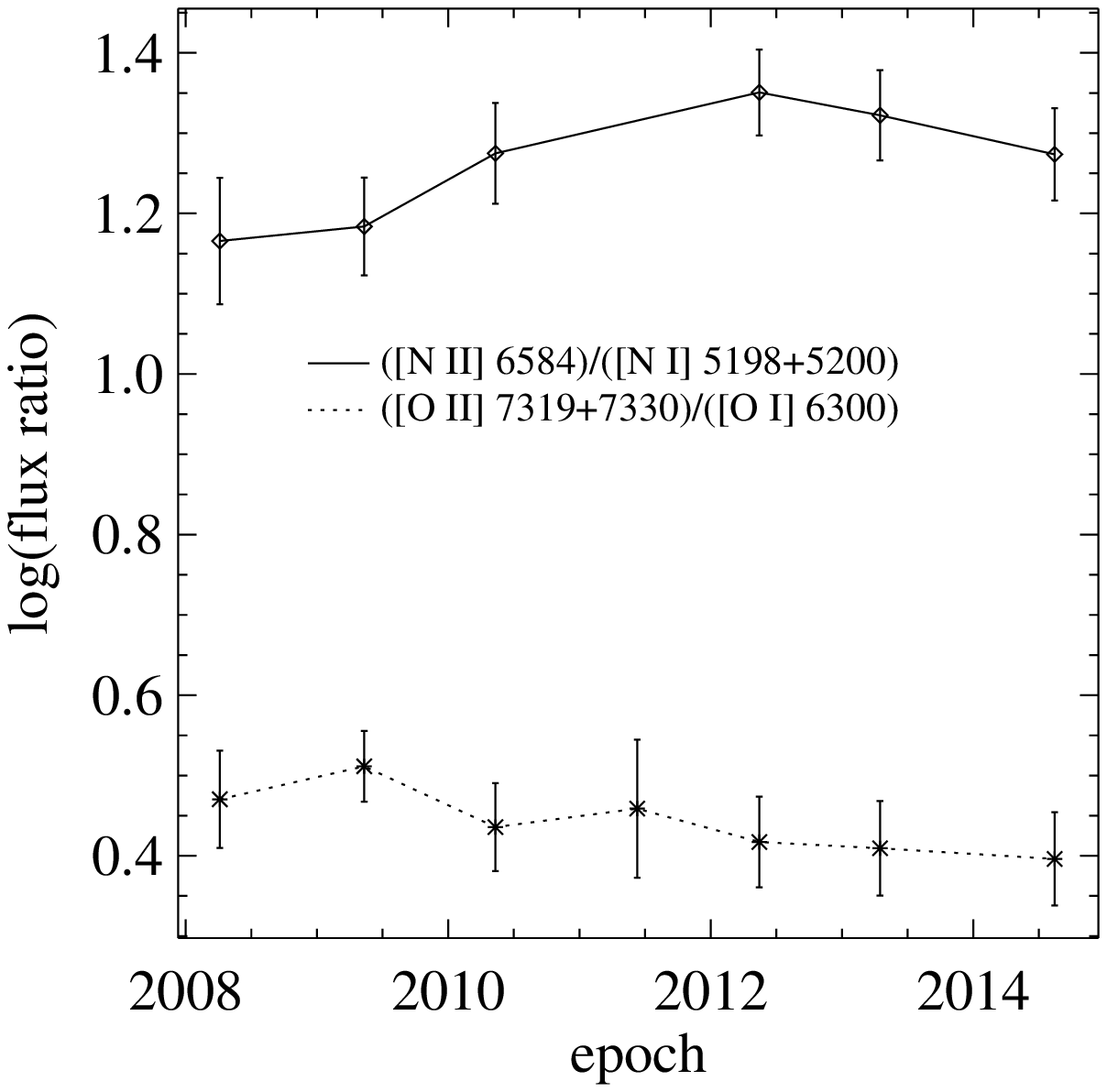}{ratio}{Left panel -- two temperature
  sensitive line ratios as a function of time. Right panel -- two line ratios
  of ionized over neutral species as a function of time.}

In Fig~\ref{ratio} we show various line ratios as a function of time. Note
that these ratios are not corrected for extinction! The left panel shows two
diagnostic ratios that are sensitive to electron temperature. Both indicate a
decrease in electron temperature since 2008, assuming that the extinction is
not increasing over time. This is inconsistent with photoionization and points
to the presence of a shock. The line ratios suggest however that the electron
temperature has been constant since 2012.

The right-hand panel of Fig.~\ref{ratio} shows line ratios of ionized over
neutral species from nitrogen and oxygen. These ratios are more difficult to
interpret. A change in density would affect the degree of ionization in the
gas, and if the density is high enough, the line ratios would also be affected
by a changing degree of collisional de-excitation. The latter doesn't appear
to be a problem as the observed [S\,{\sc ii}] line ratio indicates an electron
density well below the critical density of all lines. The gas density could be
dropping due to expansion of the nebula, or could be rising due to compression
by the shock. Both line ratios indicate a drop in the degree of ionization
(assuming that the extinction is not decreasing) at least since 2012. This is
inconsistent with photoionization and would point to either the strength of
the shock diminishing or the density of the gas increasing. Either
interpretation would point to the presence of a shock.

If the central star temperature is indeed rising, as suggested by
\citet{HJ14}, it could have caused an increase in the mass loss and wind
velocity from the central star, which is now causing shock emission in the
bipolar structure. The sudden jump in the [O\,{\sc ii}] flux, as well as the
dropping electron temperature point to a shock. The rising flux could be due
to the fact that the interaction area between wind and the bipolar structure
is still growing and is now approaching a maximum. We would like to emphasize
that the analysis of our data is not yet complete, and all scenarios presented
here are preliminary.

\acknowledgements PvH acknowledges support from the Belgian Science Policy
Office through the ESA PRODEX program.

\bibliography{proceedings}  % For BibTex

\end{document}